\documentclass{elsart}

\journal{Physica A}

\usepackage[dvips]{graphicx}
\usepackage{amsmath,amsfonts}
\usepackage[numbers,sort&compress]{natbib}

\providecommand{\av}[1]{\left\langle #1 \right\rangle}
\providecommand{\neu}[1]{\eta_{\mathrm{#1}}}

\begin{document}

\begin{frontmatter}

\title{Order-disorder transition in the Chialvo-Bak `minibrain' controlled by network geometry}

\date{6 February 2003}
\author{Joseph Wakeling}
\address{Institut de Physique Th\'eorique, Universit\'e de Fribourg P\'erolles \\ CH-1700 Fribourg, Switzerland}
\ead{joseph.wakeling@unifr.ch}
\ead[url]{http://neuro.webdrake.net/}

\begin{abstract}
We examine a simple biologically-motivated neural network, the three-layer version of the Chialvo-Bak `minibrain' [Neurosci. 90 (1999) 1137], and present numerical results which indicate that a non-equilibrium phase transition between ordered and disordered phases occurs subject to the tuning of a control parameter.  Scale-free behaviour is observed at the critical point.  Notably, the transition here is due solely to network geometry and not any noise factor.  The phase of the network is thus a design parameter which can be tuned.  The phases are determined by differing levels of interference between active paths in the network and the consequent accidental destruction of good paths.
\end{abstract}

\begin{keyword}
Phase transitions \sep Neural networks \sep Neuroscience
\PACS 05.70.Fh \sep 84.35.+i \sep 87.19.La
\end{keyword}

\end{frontmatter}

\section{Introduction}

Biological neuronal systems are certainly complex---but the source of this complexity still remains unclear.  Bearing in mind the many examples of complex behaviour arising from simple rules~\cite{KauffmanOoO,BCG_WWMP,BTW87,BTW88,BS93,BakHNW}, a number of authors~\cite{BakHNW,HebbOoB,B85,BJ87,MAJ91,AS95,SB95,CB99,BC01} have suggested that some such basic mechanics might be responsible, and a number of neural models have been proposed~\cite{B85,BJ87,MAJ91,AS95,SB95,CB99,BC01} which aim to show that even a very simple interpretation of biological neural dynamics can nevertheless produce complex patterns of learning behaviour.  Perhaps the most appealing is that created by Chialvo and Bak~\cite{CB99,BC01}: rather cheekily dubbed the `minibrain'~\cite{WB01}, it is notable for its combined use of extremal (`winner-takes-all') dynamics~\cite{BS93,KohonenSOM} and solely negative feedback to create a fast and highly adaptive learning system.

In this paper we examine the learning patterns of the simplest version of this model and present numerical results which indicate that, subject to tuning of a control parameter, a non-equilibrium transition is observed in the learning dynamics between ordered and disordered phases, similar to that observed in traditional Hopfield neural networks~\cite{H82,AGS85,G87,STS90,BHS92,SST92,WRB93}; scale-free behaviour is observed at the critical point.  Notably, the control parameter in the minibrain network is the network geometry: the network phase is therefore \emph{a design parameter of the system}.  Although a complete analytic understanding of this phenomenon is still to be developed, the underlying mechanism behind this behaviour is identified as being interference between active paths in the network.  The behaviour at the critical point can also be compared to Kohonen's self-organizing map (SOM)~\cite{KohonenSOM}, where a small stimulus can also lead to scale-free modification of the network structure~\cite{F01}.

\section{\label{sec:2}The Chialvo-Bak `minibrain'}

\begin{figure}[t]
\centerline{\includegraphics[width=0.7\textwidth]{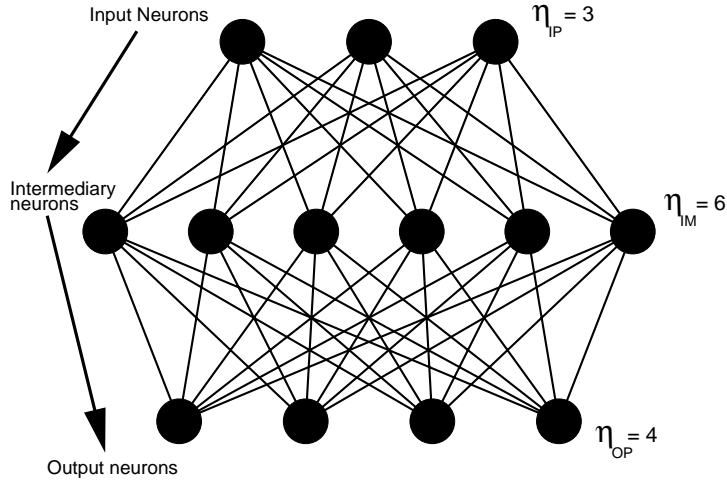}}
\caption{\label{fig:1}A three-layer minibrain network with $\neu{IP} = 3$ input neurons, $\neu{IM} = 6$ intermediary neurons and $\neu{OP} = 4$ output neurons.  Each input neuron has a (one-way) synaptic connection to every intermediary neuron, and each intermediary neuron has a similar connection to every output neuron.  Using the shorthand of Equation~(\ref{eq:1}), this network would be denoted by $\Gamma (3,6,4)$.}
\end{figure}

Although a variety of alternative network topologies are possible~\cite{CB99,BC01}, the simplest version of the minibrain consists of three separate layers of neurons (see Figure~\ref{fig:1}): a layer of $\neu{IP}$ inputs, a `hidden' intermediary layer of size $\neu{IM}$, and a layer of $\neu{OP}$ output neurons.  Each input neuron has a (one-way) synaptic connection to every intermediary neuron, and similarly each intermediary neuron is connected to every output neuron; we will use the terminology
\begin{equation}
\label{eq:1}
\Gamma\equiv\Gamma (\neu{IP},\neu{IM},\neu{OP})
\end{equation}
to denote such a network geometry.

Each synaptic connection is assigned a strength value, initially determined randomly: when an input neuron is stimulated, a signal is sent along the single strongest connection leading from it, and the intermediary neuron at the end of that connection then fires.  In the same manner, the output neuron that fires is the one at the end of the strongest connection leading from the firing intermediary neuron.  These \emph{extremal} (`winner-takes-all') dynamics, inspired both by earlier SOC models~\cite{BS93} and the SOM~\cite{KohonenSOM}, mean that we can identify an \emph{active level} of synaptic connections, the strongest connections leading from each neuron, that are collectively responsible for all decisions taken by the system.  As we will see later, the active level is an extremely important aspect of the minibrain network.

Chialvo and Bak assume that synaptic activity leaves behind a tag of some sort (biologically, perhaps a chemical trace) which identifies that synapse as `used'; feedback then takes the form of a global signal sent to the system (perhaps the release of some hormone) which reacts only with the tagged synapses.  Following the lead of Barto, Jordan et al.~\cite{B85,BJ87,MAJ91}, this global signal is extremely simple, indicating no more than whether the action taken has failed or succeeded.  What next?  If success is indicated, no further action is taken.  \emph{There is no strengthening of `good' synapses.}  If on the other hand the feedback signal is one of failure, the tagged synapses are `punished' by having their strengths reduced by a random amount.

This selectionist interpretation of biological learning dynamics may be controversial because, in particular, it clashes with the traditional Hebbian learning~\cite{HebbOoB}, where good connections are strengthened---and this has traditionally had prime place among theories of biological learning.  Chialvo and Bak cite two main reasons for their alternative viewpoint~\cite{CB99,BC01}: firstly, \emph{reinforcement of connections that produce good responses leads to addiction} unless some mechanism exists to stop the reinforcement once a correct response has been learned~\cite{P86}, whereas punishment of connections that produce bad responses by definition stops as soon as a good response is achieved; secondly, \emph{learning by mistakes is quicker}, as a system in an unknown situation is more likely to make mistakes than get things right.  That selectionism and negative feedback may in fact be the primary biological learning mechanism has also been emphasized by a number of other authors in works both experimental~\cite{Y65,FSP97} and theoretical~\cite{AlbusBBR,EdelmanND}.  A detailed account of the biological motivation for the model can be found in~\cite{CB99,BC01}.

The minibrain has already been shown to have interesting learning behaviour~\cite{CB99,BC01,WB01}, in particular being able to learn nonlinear functions such as the \textsc{xor} problem and its generalization, the parity problem.  Also, and importantly, it is a highly \emph{adaptive} model, able to quickly unlearn patterns of behaviour that are no longer successful---a direct result of the negative-only feedback~\cite{CB99,BC01,AVM00}.

\section{Control parameter for learning and order-disorder phase transition}

\begin{figure}[t]
\centerline{\includegraphics[width=0.7\textwidth]{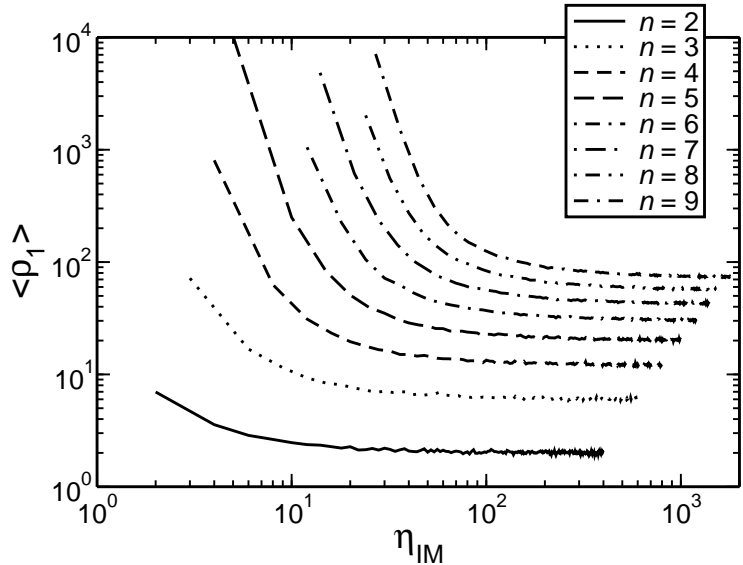}}
\caption{\label{fig:2}Mean number, $\av{\rho_{1}}$, of applications of negative feedback required to learn an input-output map of size $n$, plotted as a function of the number $\neu{IM}$ of intermediary neurons.  Data is averaged over 2048 realizations.}
\end{figure}

Arguably the simplest task one can set such a system is the learning of an input-output map (e.g., $1\to 1$, $2\to 2$, \ldots, $n\to n$): sequentially stimulating each input neuron and, if the output is unsatisfactory, applying negative feedback as described above.  Chialvo and Bak showed, although without explicit quantification, that learning time for such a map (defined as the number of times the network has to run through a complete cycle of inputs before complete success is achieved) decreases as the number $\neu{IM}$ of intermediary neurons increases~\cite{CB99}.  However, `learning time' cannot really be thought of as a physical quantity within the minibrain network itself, and hence in this paper we will use an alternative (though related) measure of learning efficiency: the total number of times, $\rho_{1}$, that negative feedback must be applied to the system---or equivalently, the number of errors made---before we can run through the complete sequence of inputs without any mistakes.  This measure can be thought of as analogous to the measure of neural network learning efficiency used in early work on the subject~\cite{H82,AGS85,G87,STS90,BHS92,SST92,WRB93} using the traditional Hopfield neural networks: the number of examples the network must see in order to learn a general function.

\begin{figure}[t]
\centerline{\includegraphics[width=0.7\textwidth]{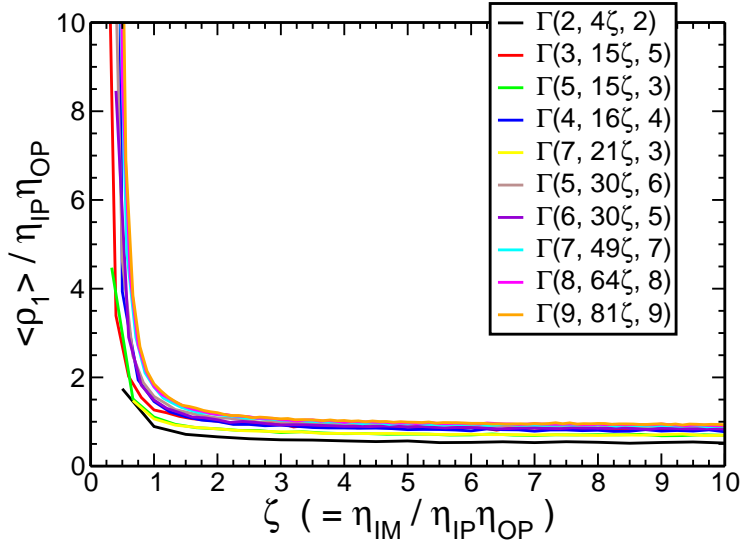}}
\caption{\label{fig:3}Mean number, $\av{\rho_{1}}$, of applications of negative feedback (in $\neu{IP}\neu{OP}$ unit) required to learn input-output maps of varying sizes, plotted as a function of control parameter $\zeta = \neu{IM}/\neu{IP}\neu{OP}$.  Note the data collapse as a result of this normalization, and the sharp transition between large and small (asymptotic minimum) values that occurs at $\zeta \approx 1$.  The data legend uses the notation of Equation~(\ref{eq:1}).  Data is averaged over 2048 realizations.}
\end{figure}

Fig.~\ref{fig:2} shows that in learning an input-output map of the above type (with $\neu{IP} = \neu{OP} = n$), the mean number, $\av{\rho_{1}}$, of applications of negative feedback required scales qualitatively in a similar manner to Chialvo and Bak's learning time: as the number of intermediary neurons $\neu{IM}$ increases, $\av{\rho_{1}}$ decreases towards an eventual asymptotic minimum whose value appears to depend on $n$.  This behaviour is quantified precisely in Fig.~\ref{fig:3}, where we investigate the effects of unequal mappings where $\neu{IP}\neq\neu{OP}$, e.g., $i\to i\bmod\neu{OP}$: both the minimum value of $\av{\rho_{1}}$ and the decrease in negative feedback that occurs with increasing $\neu{IM}$ collapse when $\av{\rho_{1}}$ and $\neu{IM}$ are normalized with respect to $\neu{IP}\neu{OP}$, and hence we can define a control parameter for learning time,
\begin{equation}
\label{eq:2}
\zeta = \frac{\neu{IM}}{\neu{IP}\neu{OP}}
\end{equation}

\begin{figure}[t]
\centerline{\includegraphics[width=0.7\textwidth]{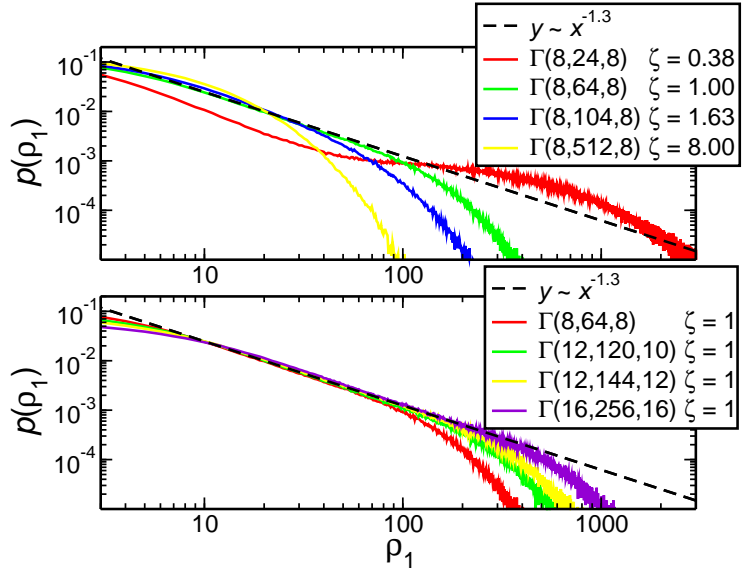}}
\caption{\label{fig:4}Probability function $p(\rho_{1})$ of the number of applications of negative feedback required to adapt to a new input-output map, based on the control parameter $\zeta$: (a)~different distributions of learning time for a network with fixed $\neu{IP} = \neu{OP} = 8$ and varying $\neu{IM}$, with power-law behaviour at the critical point $\zeta = \neu{IM}/\neu{IP}\neu{OP} = 1$; (b)~power-law (scale-free) behaviour at the critical point for different system sizes, with $p(\rho_{1})\sim \rho_{1}^{-1.3}$.  Data taken over $10^{6}$ adaptions.}
\end{figure}

It can be seen that at $\zeta \approx 1$ a sharp transition occurs in $\av{\rho_{1}}$ between the asymptotic minimum and truly immense values.  Numerical results (Fig.~\ref{fig:4}) suggest that this is a genuine phase transition: if once the minibrain has learned a given input-output mapping, we randomly reassign a new output to one of the inputs, and repeat this process, we find that the probability function of negative feedback $\rho_{1}$ required to adapt to the new map varies between an ordered phase ($\zeta > 1$) where there is a maximum cutoff to learning times, with $\rho_{1}$ distributed exponentially, and a disordered phase ($\zeta < 1$) where learning is almost impossible, with $\rho_{1}$ strongly biased towards huge values.  The probability function of $\rho_{1}$ becomes scale-free at the critical point $\zeta = 1$, scaling as a power law with exponent $-1.3$.

To understand what this means physically for the network, we have to recall what the meaning of negative feedback is.  Decreasing the strength of a synaptic connection almost certainly removes it from the active level of synapses; another connection replaces it.  A large number of negative feedback signals therefore corresponds to a large change in the geometry of the active level of the network---an `avalanche' in the network landscape, so to speak.  But what is it that causes such large avalanches?

The answer lies in \emph{path interference}, the \emph{overlapping of the active paths} leading from two or more distinct inputs.  In the three-layer minibrain, this is as simple as the strongest connections from two different inputs pointing to the same intermediary neuron.  Broadly speaking, if the paths from two or more neurons overlap in this fashion, it is likely that (at least) one of the inputs is connected to the wrong output neuron; we can make this explicitly true if we require that the input-output map to be learned be one-to-one.  The negative feedback, however, will affect \emph{both} paths by removing the intermediary-output connection from the active level.  Thus, a correct wiring of input to output can be destroyed by such interference.

A brief statistical analysis reveals the behaviour of this factor.  Suppose we have an arbitrary minibrain network, $\Gamma (\neu{IP},\neu{IM},\neu{OP})$.  Labeling the input neurons by $0, 1, 2, \dots , \neu{IP}-1$, let us define, for a given input $i$, $\mu (i)$ to be the intermediary neuron led to by the strongest synaptic connection from $i$, and let $p_{i}^{\Gamma}$ be the probability that at least two input neurons in the set $\{ 0, \dots , i\}$ share the same value of $\mu$.  Thus, the probability of having path interference in the network is given by $P_{\Gamma} = p_{\neu{IP}-1}^{\Gamma}$.  Assuming that the values of $\mu$ are randomly distributed, it follows that $p_{0}^{\Gamma} = 0$ and $p_{1}^{\Gamma} = 1/\neu{IM}$, and more generally (assuming $\neu{IM}\geq\neu{IP}$, since obviously $P_{\Gamma}=1$ if this is not the case),
\begin{equation}
p_{i + 1}^{\Gamma} = p_{i}^{\Gamma} + \frac{i + 1}{\neu{IM}}(1 - p_{i}^{\Gamma})
\end{equation}

Now, substituting for $\neu{IM}$ in terms of the control parameter $\zeta$, we have:
\begin{equation}
\label{eq:3}
p_{i + 1}^{\Gamma} = p_{i}^{\Gamma} + \frac{i + 1}{\neu{IP}\neu{OP}\zeta}(1 - p_{i}^{\Gamma})
\end{equation}

From which we can calculate,
\begin{equation}
\label{eq:4}
p_{n}^{\Gamma} = \sum_{i=1}^{n} \left[ {n \choose i}(-1)^{i-1} \left(\neu{IP}\neu{OP}\right)^{-i} \zeta^{-i} \right]
\end{equation}

\begin{figure}[t]
\centerline{\includegraphics[width=0.7\textwidth]{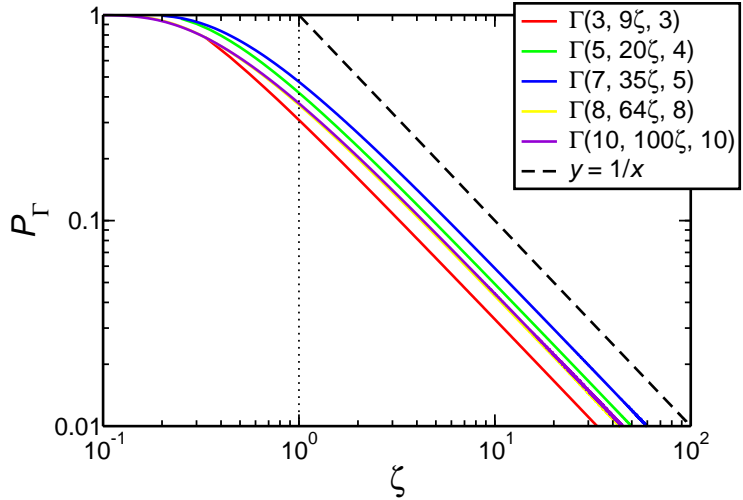}}
\caption{\label{fig:5}Probability $P_{\Gamma}$ of finding path interference in various networks $\Gamma (\neu{IP},\neu{IM},\neu{OP})$, plotted as a function of control parameter $\zeta$.  Note the transition at $\zeta\approx 1$ between exponential and power-law behaviour.  Values numerically calculated from Equation~(\ref{eq:3}).}
\end{figure}

Assuming large, fixed $\neu{IP}$ and $\neu{OP}$, the higher-order terms in Equation~(\ref{eq:4}) will vanish for $\zeta > 1$, and so $P_{\Gamma}$ will scale approximately according to a power law of $\zeta$, with exponent $-1$.  Conversely, if $\zeta < 1$ then the higher-order terms are brought back into play and $P_{\Gamma}$ scales approximately exponentially; Fig.~\ref{fig:5} shows nicely how the critical point $\zeta = 1$ acts as a transition point between these two scalings.  The phase transition can thus be seen as a transition between a high-interference phase where learned patterns are frequently destroyed by accident as a result of mistakes elsewhere, and a low-interference phase where active connections remain largely stable.

It is interesting to compare this behaviour with the learning patterns observed in more traditional neural models, for example the large body of work on Hopfield networks~\cite{H82,AGS85,G87,STS90,BHS92,SST92,WRB93}.  Typically the performance of an $N$-input network in learning a function is studied with respect to the number $P$ of examples presented to it (analogous to the parameter $\rho_{1}$ in this work, which is, recall, the number of times the minibrain must be shown where it is wrong before the error rate decreases to zero).  A transition between ordered and disordered phases, similar to that observed here, was observed by Amit et al.~\cite{AGS85}, occurring as a result of the temperature of the system; however, the behaviour we see in the minibrain is more reminiscent of the behaviour observed by Gardner~\cite{G87}, who showed that, for fixed temperature, the phase transition in the Hopfield model occurred subject to the connectivity of neurons.  In the minibrain, the path interference driving the phase transition is similarly a result solely of the network geometry---but in addition the minibrain is noise- (temperature-) resistant~\cite{BC01}.  The network phase can therefore be seen as robust with respect to the environment.

\section{Conclusions}

While only a very simple interpretation of real neural dynamics, we have seen that the minibrain displays an interesting variety of behaviour, with a non-equilibrium transition between ordered and disordered phases occurring solely on the basis of network geometry.  Thus \emph{the network phase is not due to external noise but is instead a design component}.  The system can be tuned---for instance, by evolution---to the most advantageous phase.

The question of \emph{which} phase is most advantageous, of course, depends on the minibrain's surrounding environment.  This immediately motivates further research into the question of which environments might favour which of the different phases.  In addition, considering the biological motivation for the model, one might want to investigate whether behaviour resembling any (or all) of these phases is to be found in real neural systems of different types.

The results here also emphasize the importance of negative feedback as providing behavioural control in learning systems.  The accidental destruction of good connections that drives the different phases in the minibrain model relies on the fact that, because of the negative-only feedback, the active and inactive connections in the network are of similar strength~\cite{CB99,BC01,AVM00}.  Not only are such systems faster and more adaptive~\cite{CB99,BC01}, they also place in the hands of the designer aspects which would otherwise be in the hands of the environment.

In summary, the minibrain's combination of negative-only feedback and extremal dynamics can be seen to cause complex behavioural patterns that depend much more on internal characteristics of the network than any external pressures.  This simple neural model has already been shown to have interesting implications for learning systems and the importance of selectionist procedures; the results presented here extend this to the wider context of phase transitions on networks and their origins.

\section*{Acknowledgments}

Special thanks and many fond memories to Per Bak, who took the time despite serious illness to offer encouragement and many helpful suggestions.  Grateful thanks also to Maya Paczuski, Ginestra Bianconi, Andrea Capocci, Damien Challet, Dante Chialvo, Paulo Laureti, and Yi-Cheng Zhang, for advice and stimulating conversations, and to anonymous referees for their comments.  This work was supported in part by Swiss National Science Foundation grant no.\ 20-61470.00.

\bibliography{jrw_journals,phase-mb}

\begin{thebibliography}{10}
\expandafter\ifx\csname url\endcsname\relax
  \def\url#1{\texttt{#1}}\fi
\expandafter\ifx\csname urlprefix\endcsname\relax\def\urlprefix{URL }\fi

\bibitem{KauffmanOoO}
S.~Kauffman, The Origins of Order: Self-Organization and Selection in
  Evolution, Oxford University Press, Oxford, 1993.

\bibitem{BCG_WWMP}
E.~R. Berlekamp, J.~H. Conway, R.~K. Guy, Winning Ways For Your Mathematical
  Plays, A.\ K.\ Peters, Ltd., 2001.

\bibitem{BTW87}
P.~Bak, C.~Tang, K.~Wiesenfeld, Phys.\ Rev.\ Lett. 59 (1987) 381.

\bibitem{BTW88}
P.~Bak, C.~Tang, K.~Wiesenfeld, Phys.\ Rev.\ A 38.

\bibitem{BS93}
P.~Bak, K.~Sneppen, Phys.\ Rev.\ Lett. 71 (1993) 4083.

\bibitem{BakHNW}
P.~Bak, How Nature Works: The Science of Self-Organized Criticality, Oxford
  University Press, Oxford, 1997.

\bibitem{HebbOoB}
D.~O. Hebb, The Organization of Behaviour, Wiley, New York, 1949.

\bibitem{B85}
A.~G. Barto, Hum.\ Neurobiol. 4 (1985) 229.

\bibitem{BJ87}
A.~G. Barto, M.~I. Jordan, Proc.\ IEEE Int.\ Conf.\ on Neural Networks 2 (1987)
  629.

\bibitem{MAJ91}
P.~Mazzoni, R.~A. Andersen, M.~I. Jordan, Proc.\ Natl.\ Acad.\ Sci.\ USA 88
  (1991) 4433.

\bibitem{AS95}
P.~Alstr{\o}m, D.~Stassinopoulos, Phys.\ Rev.\ E 51 (1995) 5027.

\bibitem{SB95}
D.~Stassinopoulos, P.~Bak, Phys.\ Rev.\ E 51 (1995) 5033.

\bibitem{CB99}
D.~R. Chialvo, P.~Bak, Neurosci. 90 (1999) 1137.

\bibitem{BC01}
P.~Bak, D.~R. Chialvo, Phys.\ Rev.\ E 63 (2001) 031912.

\bibitem{WB01}
J.~Wakeling, P.~Bak, Phys.\ Rev.\ E 64 (2001) 051920.

\bibitem{KohonenSOM}
T.~Kohonen, Self-Organizing Maps, Springer-Verlag, Berlin, 2001.

\bibitem{H82}
J.~J. Hopfield, Proc.\ Natl.\ Acad.\ Sci.\ USA 79 (1982) 2554--2558.

\bibitem{AGS85}
D.~J. Amit, H.~Gutfreund, H.~Sompolinsky, Phys.\ Rev.\ Lett. 55 (1985) 1530.

\bibitem{G87}
E.~Gardner, J.\ Phys.\ A 20 (1987) 3453.

\bibitem{STS90}
H.~Sompolinsky, N.~Tishby, H.~S. Seung, Phys.\ Rev.\ Lett. 65 (1990) 1683.

\bibitem{BHS92}
E.~Barkai, D.~Hansel, H.~Sompolinsky, Phys.\ Rev.\ A 45 (1992) 4146.

\bibitem{SST92}
H.~S. Seung, H.~Sompolinsky, N.~Tishby, Phys.\ Rev.\ A 45 (1992) 6056.

\bibitem{WRB93}
T.~L.~H. Watkin, A.~Rau, M.~Biehl, Rev.\ Mod.\ Phys. 65 (1993) 499.

\bibitem{F01}
J.~A. Flanagan, Phys.\ Rev.\ E 63 (2001) 036130.

\bibitem{P86}
G.~Parisi, J.\ Phys.\ A 19 (1986) L617.

\bibitem{Y65}
J.~Z. Young, Proc.\ Royal Soc.\ London B 163 (1965) 285.

\bibitem{FSP97}
R.~M. Fitzsimonds, H.~J. Song, M.~M. Poo, Nature 388 (1997) 439.

\bibitem{AlbusBBR}
J.~S. Albus, Brains, Behaviour and Robotics, BYTE Books, McGraw-Hill,
  Peterborough, NH, 1981.

\bibitem{EdelmanND}
G.~M. Edelman, Neural Darwinism: The Theory of Neuronal Group Selection, Basic
  Books, New York, 1987.

\bibitem{AVM00}
T.~Araujo, R.~Vilela~Mendes, Complex Syst. 12 (2000) 357.

\end{thebibliography}

\end{document}